# Verification Testing of MQXFA $Nb_3Sn$ Wires Procured Under LARP

Jeremy Levitan, Jun Lu, Vito Lombardo, and Lance Cooley *Senior Member, IEEE*

***Abstract*— The High-Luminosity LHC Accelerator Upgrade Project (AUP) in the U.S. will construct quadrupole magnets to be delivered to CERN. An initial 3 tons, over 600 km total length, of conductor was procured under the LHC Accelerator R&D Program (LARP) for this project. Programs for quality control (QC) at the supplier and quality verification (QV) at the laboratories were solidified into components of the overall quality plan for strand procurement under AUP. Measurements of the critical current ($I_c$) and residual resistance ratio (RRR), and related probes and techniques, are central to the quality plan. Described below is the verification testing that has taken place at the National High Magnetic Field Laboratory (NHMFL). Testing challenges are presented by the high sensitivity of these wires. In addition, new RRR test software was developed to accommodate challenges presented by meeting international standards with existing configurations of test strands.**

## I. Introduction

The High-Luminosity LHC Accelerator Upgrade Project (AUP) in the U.S. will construct quadrupole magnets to be delivered to CERN [1]. An initial 3 tons, over 600 km total length, of conductor for this project was procured under the LHC Accelerator R&D Program (LARP). These wires must meet challenging critical current ($I_c$) and residual resistivity ratio (RRR) specifications, which are shown in Table I. In addition to quality control tests performed by the manufacturer, verification tests are performed by the National High Magnetic Field Laboratory (NHMFL). The process and results of these tests are presented and compared to the manufacturer's data below.

## II. Experimental

### A. Heat Treatment

Strands approximately 1 meter long were cut from verification sample lengths sent from Fermilab for the $I_c$ tests. The strand samples were wound with 1 kg of tension on $Ti_6Al_4V$ barrels used for heat treating and testing ITER $Nb_3Sn$ wires, with no modification of the barrels or mounting procedures from the description in [2]. Straight wires approximately 25 cm long were inserted into 1 mm interior diameter quartz tubes to create RRR samples. The ends of the $Nb_3Sn$ wire for both the $I_c$ and RRR samples were welded shut using a spot welder to prevent tin leak during the heat treatment [3], [4]. Groups of samples were wrapped in stainless steel foil before being placed in the center of the furnaces.

TABLE I
PARAMETERS FROM THE LARP $Nb_3Sn$ SPECIFICATIONS [1]

| Description | Specification |
|---|---|
| $I_c$ 4.22 K, 12 T (A) | ≥ 632 |
| $I_c$ 4.22 K, 15 T (A) | ≥ 331 |
| $n$-value 4.22 K, 15 T | > 30 |
| RRR | ≥ 150 |
| Diameter (mm) | 0.85 |
| Sub-elements | 108/127 |
| Nb:Sn ratio | 3.6:1 |

Both the $I_c$ and RRR samples were heat treated in UHP argon (Ar) gas using following schedule,

Ramp at 25 °C/h to 210 °C, soak for 48 h at 210 °C,
Ramp at 50 °C/h to 400 °C, soak for 48 h at 400 °C,
Ramp at 50 °C/h to 665 °C, soak for 75 h at 665 °C,
Furnace cool.

The cross-section of a heat treated wire is shown in Fig. 1. It should be noted that the heat treatment at the supplier is 50 hours at 665 °C, 25 hours shorter than ours.

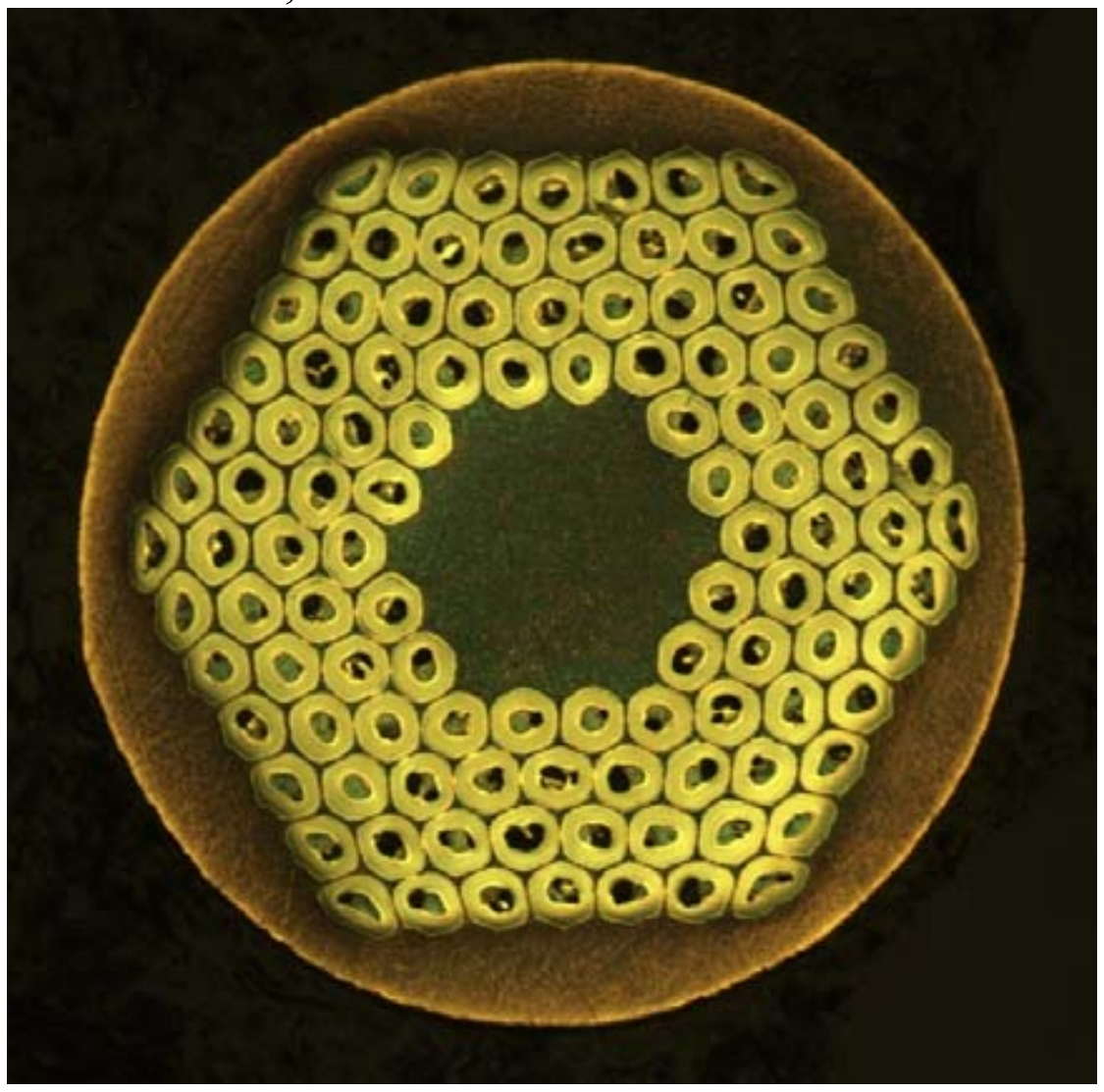

Fig. 1. Cross-section of MQXFA $Nb_3Sn$ after heat-treatment.

This work is financially supported by DOE through Fermi National Accelerator Laboratory. This work was performed at the National High Magnetic Field Laboratory, which is supported by National Science Foundation Cooperative Agreement No. DMR-1644779 and the State of Florida.

J. Levitan and J. Lu are with Magnet Science and Technology, National High Magnetic Field Laboratory, Tallahassee, FL 32310 USA, (Corresponding author: Jun Lu, junlu@magnet.fsu.edu.) L. D. Cooley is with the Applied Superconductivity Center, National High Magnetic Field Laboratory, Tallahassee, FL 32310 USA. Vito Lombardo is with Technical Division of the Fermi National Accelerator Laboratory, Batavia, IL 60510 USA.

## B. Verification Tests

### 1) Critical Current ($I_c$)

After the heat treatment, each wire was detached at one end and gently loosened from the ITER barrel. It was then tightened by applying downward pressure with a finger, which moved along the helical groove of the ITER barrel. The wire was reattached and soldered to the copper rings. This process removed approximately 3 mm of slack from the wire. The barrel was then mounted on a test probe, with pressure contacts to the copper terminals, as shown in Fig. 2. Two voltage taps with lengths of 25 and 50 cm were soldered across the center turns. The 50 cm length was used for the measurement. $I_c$ was determined by measuring voltage in a liquid helium bath as current was increased by 1 A steps in a superconducting magnet. $I_c$ was recorded at 12, 13, 14, and 15 T with a criterion of 10 μV/m. The current-voltage curves were also used to determine the transition index *n* by a power law fit for data between 10 μV/m and 100 μV/m.

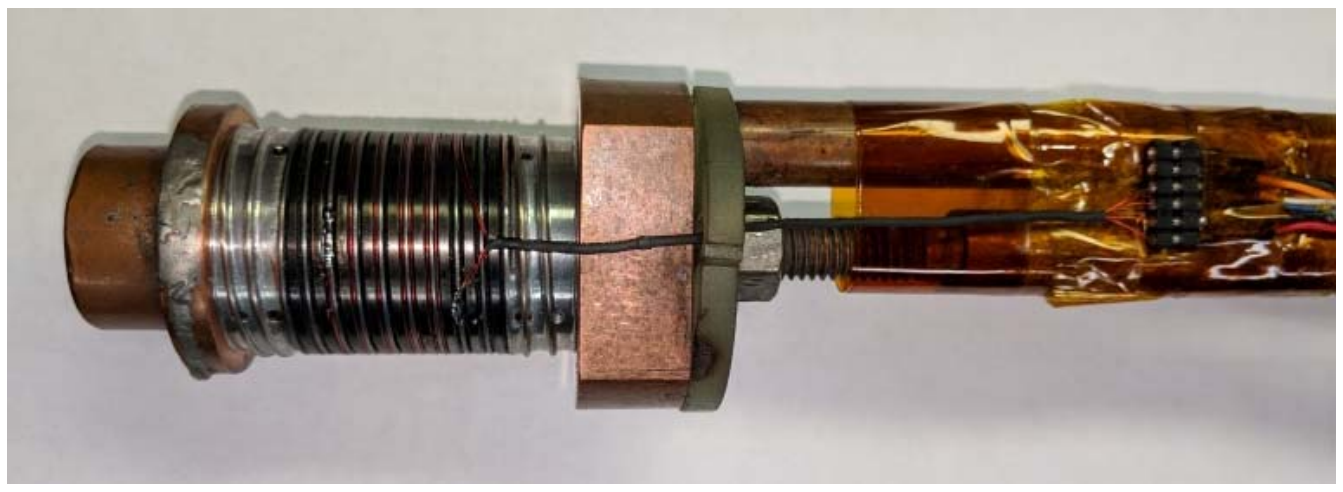

Fig. 2. A prepared $I_c$ sample attached to a probe for testing.

### 2) Residual Resistance Ratio (RRR)

The set up used for RRR measurements was the same as previously reported [3], [4]. Unlike for the ITER project, which defined RRR as the ratio of resistances at 273 K and 20 K, the LARP specification called for the use of the IEC standard [5], which is defined as the ratio of resistance at room temperature (295 K) to that determined by the intersection of two straight lines extrapolated from the resistance versus temperature curve, as shown on Fig. 3. A LabVIEW program was developed to measure resistance versus temperature during a slow warm-up from 4.2 K. The temperature was measured by a Cernox temperature sensor, and the typical warm-up rate was 10 mK per second. The resistance was obtained by measuring voltage when -1 and +1 A current was applied using a Keithley 2400 SourceMeter®. The resistance vs. temperature curve was then analyzed by a second LabVIEW program, which fitted the two sections near the normal-to-superconducting transition of the *R* vs. *T* curve with straight lines as shown in Fig. 3. The low temperature resistance used to determine RRR was then defined by the intercept of the two straight lines (point A in Fig. 3) [5]. Since the RRR measured by the manufacturer was defined as $R_{295K}/R_{20K}$, resistance data at 20 K was also obtained to calculate $R_{295K}/R_{20K}$ and compare with the manufacturer's data.

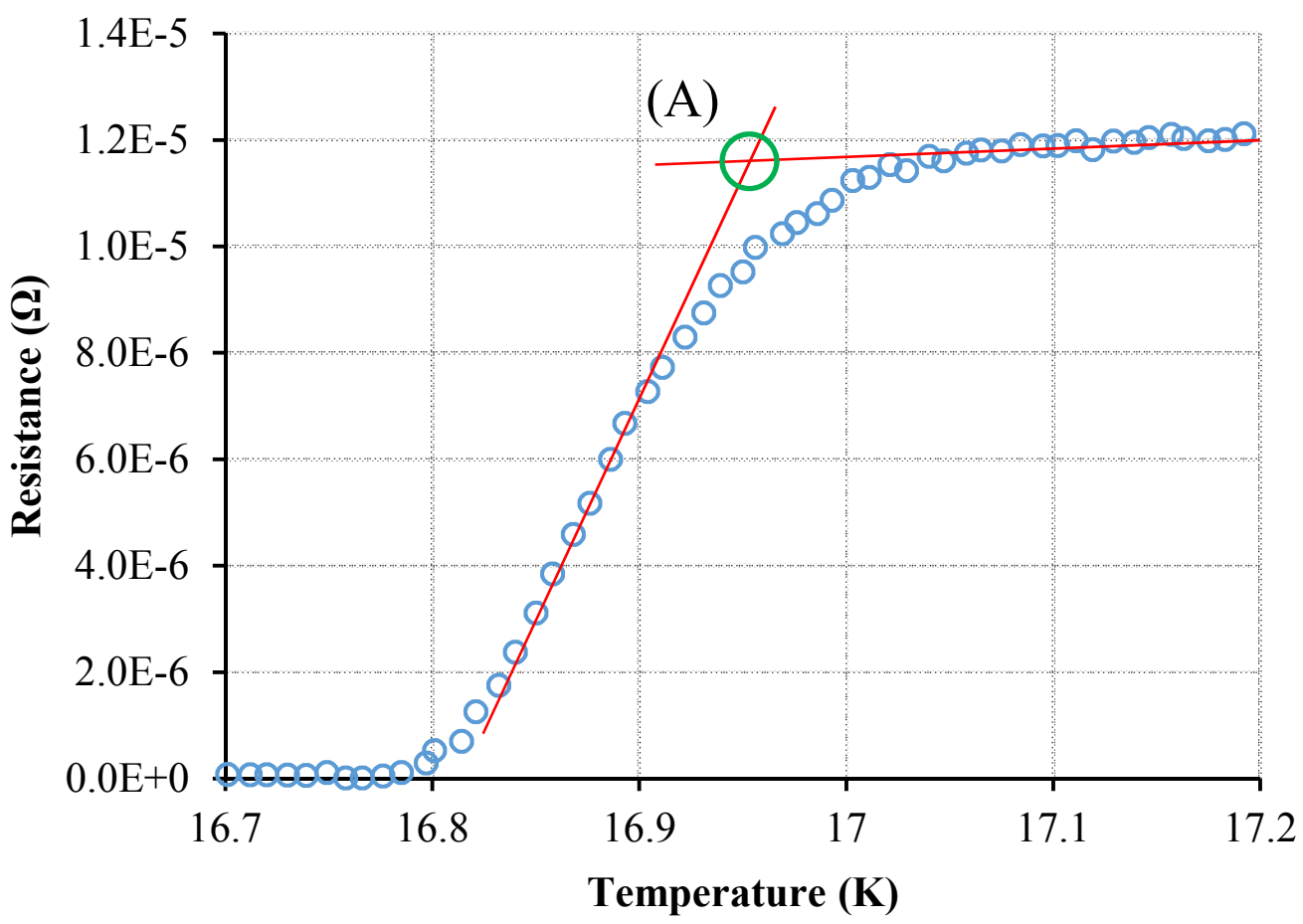

Fig. 3. A resistance versus temperature curve. RRR is obtained from the ratio of resistance at 295 K (not shown) and point (A) on the curve.

# III. Results

## A. Critical Current ($I_c$)

The final step of the supplier's heat treatment was 665 °C for 50 hours instead of 75 hours as used by the NHMFL. An investigation was performed by Fermilab and the vendor to determine the impact of the extended heat treatment. A comparison was made of the vendor data, using the average of the 187 samples measured for $I_c$ at 12, 13, 14, and 15 T. 50 samples were used for 665°C/75h and 137 were used for 665°C/50h. This resulted in a 2.9% difference for $I_c$ data obtained at 12 T, and a 4.9% difference for $I_c$ data obtained at 15 T. Results of the investigation are shown in Table II.

TABLE II
HEAT TREAMENT COMPARISON RESULTS

| | $I_c$ at 12 T | $I_c$ at 15 T | RRR |
|---|---|---|---|
| Vendor 665°C/75h average | 713 A | 388 A | 300 |
| Vendor 665°C/50h average | 693 A | 370 A | 301 |
| Vendor 665°C/75h : 665°C/50h | 102.9% | 104.9% | 99.7% |
| Lab 665°C/75h : Vendor 665°C/75h | 100.3% | 102.1% | N/A |
| Lab 665°C/75h : Vendor 665°C/50h | 103.1% | 107.1% | N/A |

The average and standard deviation of our results are summarized in Table III. Fig. 4 shows our $I_c$ results at 12 and 15 T after they have been corrected for the difference in heat treatments between the NHMFL and the supplier. The corresponding *n*-values at 15 T are shown in Fig. 5. The results measured by the supplier are also plotted in each figure for comparison.

TABLE III
SUMMARY OF RESULTS

| | NHMFL | | Supplier | |
|---|---|---|---|---|
| | Mean | σ | Mean | σ |
| $I_c$ 12T (A) | 712.7 | 16.5 | 694.9 | 14.3 |
| $I_c$ 13T (A) | 593.1 | 13.2 | 572.8 | 11.5 |
| $I_c$ 14T (A) | 488.1 | 9.8 | 465.6 | 9.7 |
| $I_c$ 15T (A) | 395.5 | 8.5 | 371.2 | 8.5 |
| *n* 12T | 43.8 | 2.8 | 44.3 | 4.4 |
| *n* 15T | 38.4 | 2.6 | 38.3 | 5.8 |
| RRR-IEC | 375.1 | 69.8 | N/A | N/A |
| $R_{295K}/R_{20K}$ | 323.7 | 55.0 | 295.8 | 46.2 |

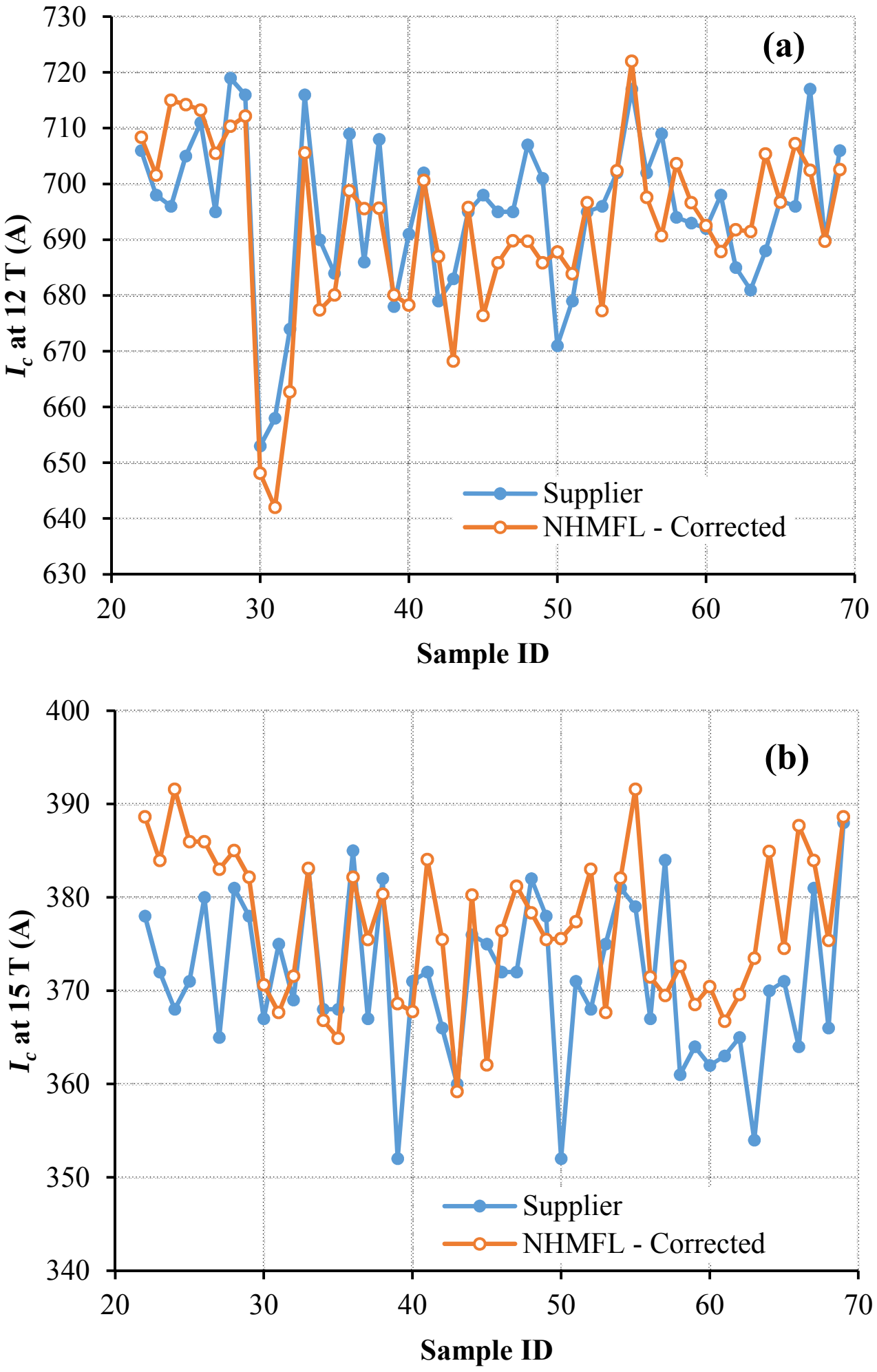

Fig. 4. $I_c$ data comparison between the corrected NHMFL and supplier data at (a) 12 T and (b) 15 T. The NHMFL results have been adjusted by 2.8% and 4.7% for 12 and 15 T respectively to correct for the difference in heat treatments.

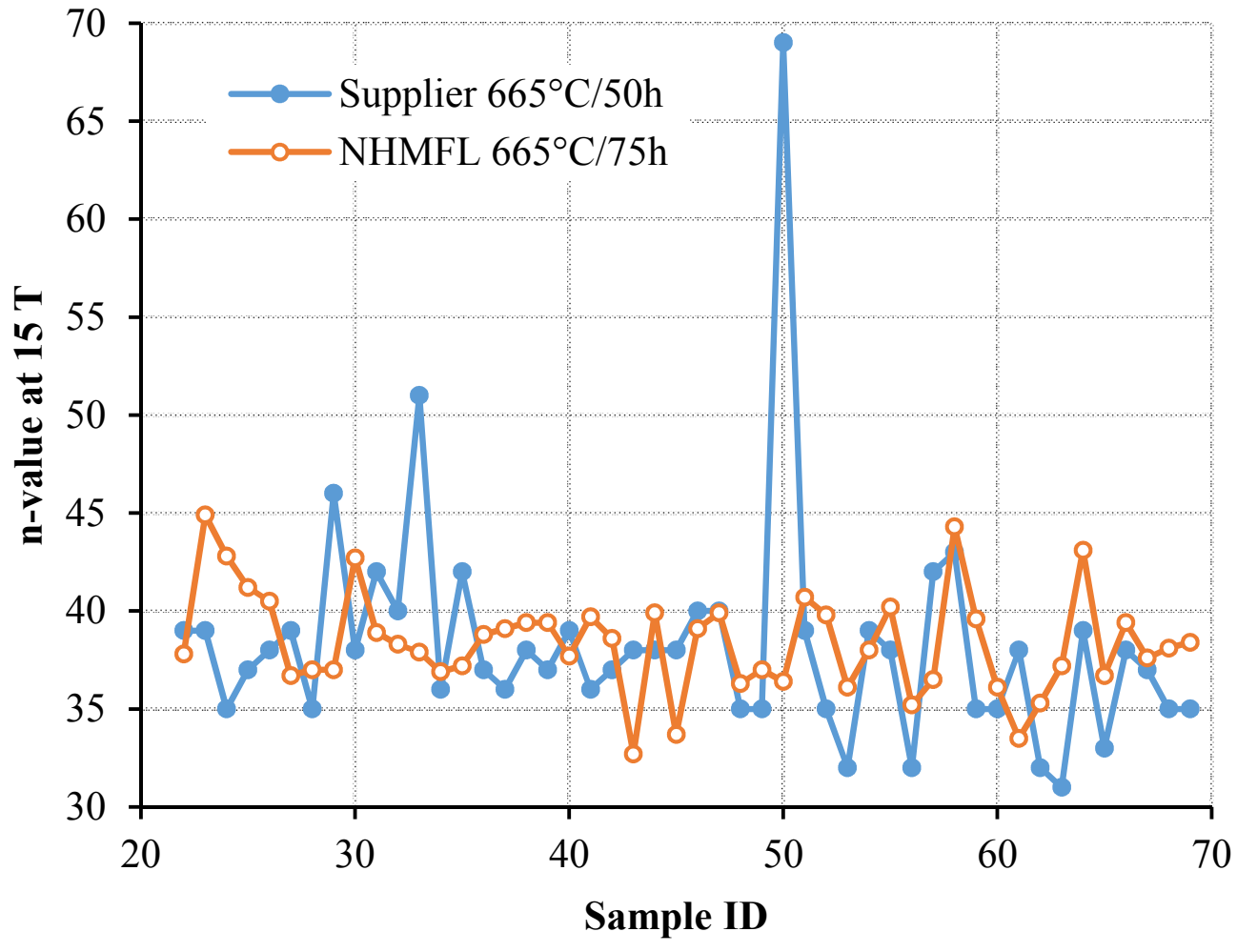

Fig. 5. *n*-value at 15 T. Comparison between the NHFML and supplier.

## B. *Residual Resistance Ratio (RRR)*

Fig. 6 shows RRR defined by the IEC standard [5] and $R_{295K}/R_{20K}$ along with the supplier's $R_{295K}/R_{20K}$ data for comparison. The average and standard deviations of each set of data are presented in Table III. From the results, the average RRR-IEC is about 16% higher than $R_{295K}/R_{20K}$ due to the fact that the IEC definition calls for resistance near the normal-to-superconducting transition, which is at about 17 K for these samples, about 3 K lower than the 20 K method.

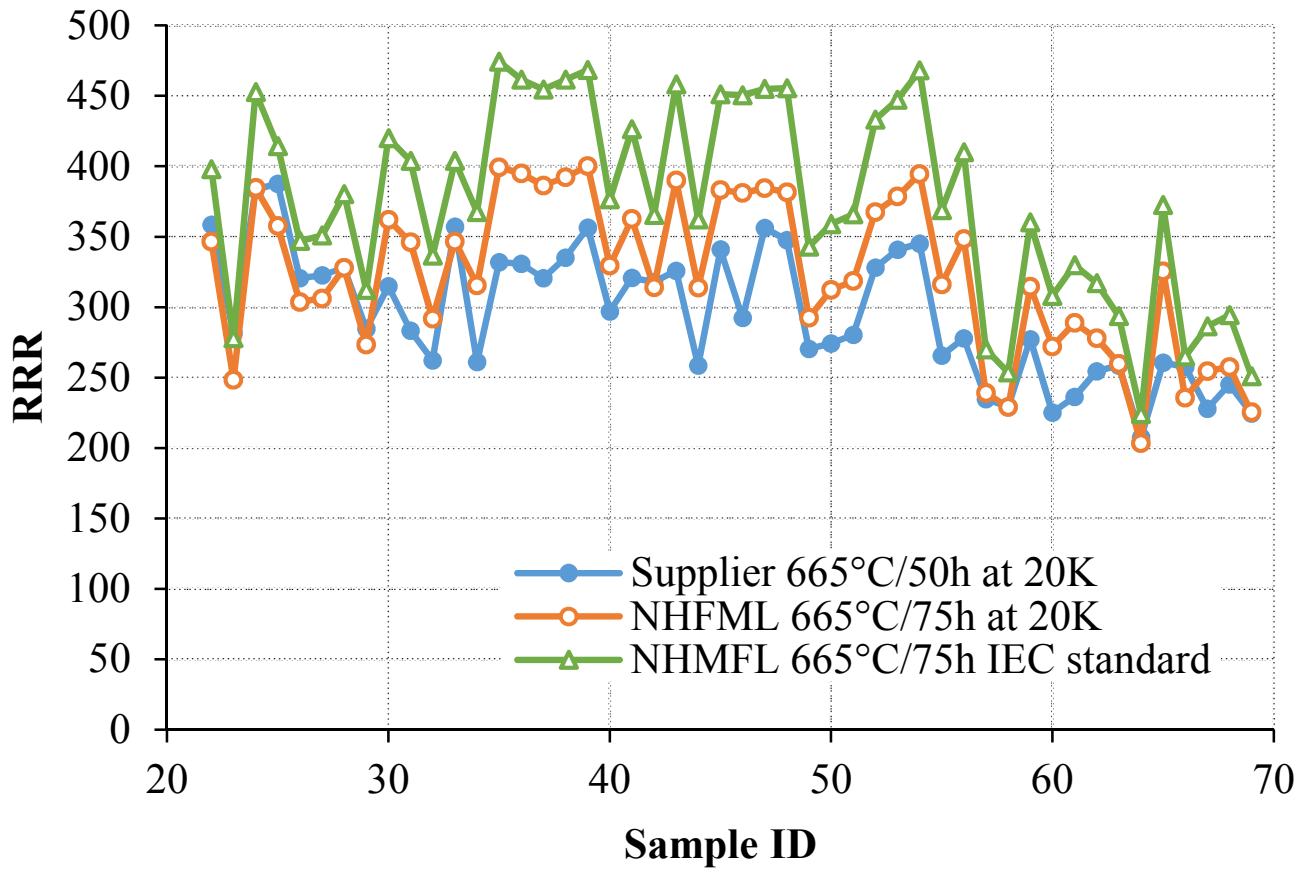

Fig. 6. RRR-IEC and $R_{295K}/R_{20K}$ data measured by the NHMFL and $R_{295K}/R_{20K}$ data measured by the supplier.

# IV. Discussion

## A. *Critical Current ($I_c$)*

It was observed that after the heat treatment, the wire was loose on the ITER barrel due to the thermal contraction mismatch between the wire and the barrel during the cool down to room temperature. This looseness under electromagnetic force during the testing made the sample prone to premature quenches. Therefore, it is reasonable to expect that tightening the wire to the ITER barrel after heat treatment would reduce the probability of premature sample quench. Our previous experience on ITER internal-tin $Nb_3Sn$ wire, however, indicated that tightening ITER $Nb_3Sn$ wire to the barrel causes more frequent premature sample quenches.

In order to study the effect of the wire tightening process, we experimented with and without the tightening process and the results are shown in Fig. 7. The label 'tightened hot' means that after the 'not tightened' sample was soldered to the copper terminal rings and tested, an attempt was made to tighten the sample on the same barrel while the solder was melted (hot). The label 'tightened' refers to new samples that were tightened before soldered to the copper rings. These were created after the success of the 'tightened hot' samples. From Fig. 7, it is evident that the tightening process produces significantly more consistent and higher $I_c$.

Sample 9 was the only sample that did not increase between 'not tightened' and 'tightened hot'. The sample quenched prematurely at 601 A which was most likely caused from damage during the tightening process. The risk of this damage was reduced greatly by tightening the wire before soldering to the barrel in the 'tightened' samples.

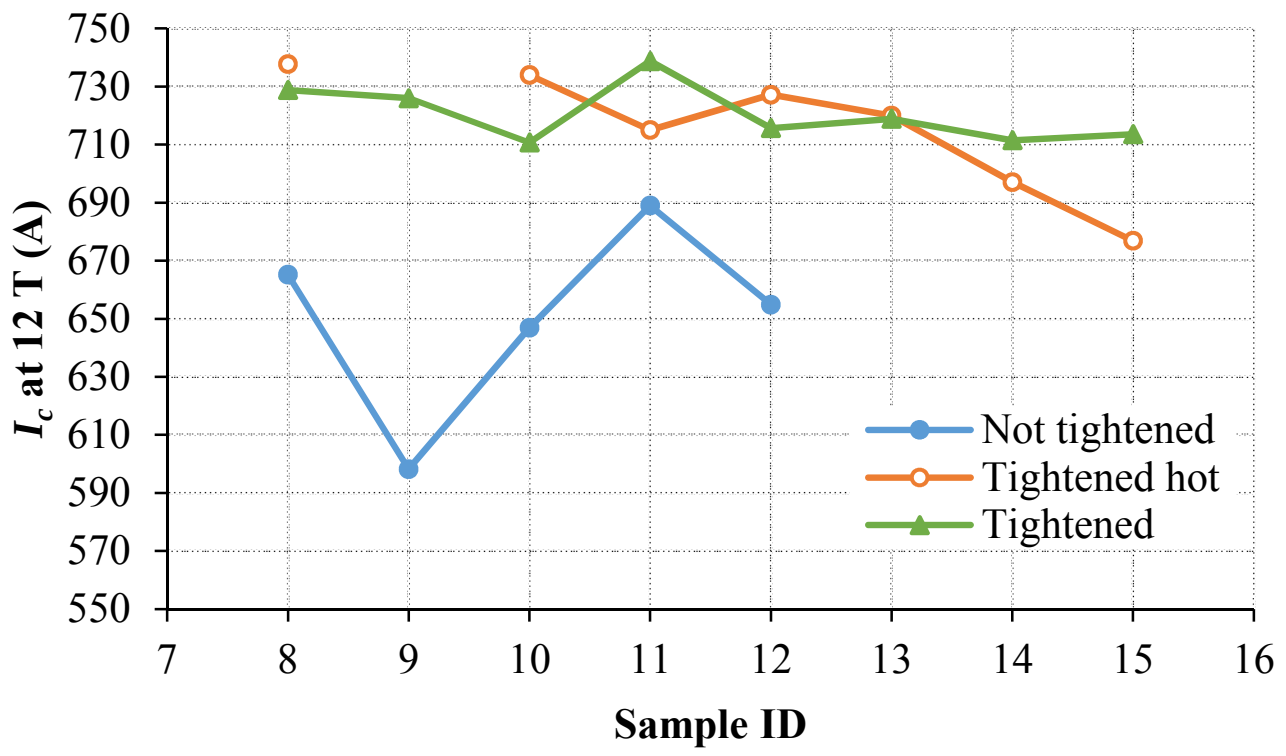

Fig. 7. $I_c$ at 12 T for samples not tightened, tightened at soldering temperature after initial soldering to Cu terminals, and tightened before soldering.

There is an obvious difference between the LARP wire and ITER wire in response to the tightening process. While ITER wires are very sensitive to damage by the tightening process, LARP wires can tolerate some handling (slightly bending) such as the tightening process. It is speculated that the ITER wire design of a single diffusion barrier might be responsible for its sensitivity to the tightening process (slightly bending). LARP wires with distributed diffusion barriers for each sub-element makes the diameter of the non-copper region much smaller than that of ITER wires. With soft copper between the sub-elements, the LARP samples are more forgiving to some post-heat-treatment handling. The higher $I_c$ values from the tightened samples can be explained by strain sensitivity of $Nb_3Sn$ where compressive intrinsic thermal strain is reduced as a result of the tightening process.

Our systematically higher $I_c$ than that of the supplier can be explained by our 25 hour longer heat treatment time during the 665 °C stage. This could lead to more thorough Nb and Sn reaction or partial reaction of Nb diffusion barriers to form $Nb_3Sn$ and, therefore, to higher $I_c$ [6].

### *B. Residual Resistance Ratio (RRR)*

Despite the fact that our heat treatment time is 25 hours longer than that of the manufacturer, which could lead to lower RRR as seen in Table II, the $R_{295K}/R_{20K}$ by NHMFL is slightly higher than that of the supplier (Fig. 6). Since a previous benchmarking by the two labs using the same set RRR sample showed very good agreement, this slight but systematic difference might be attributed to slight differences in the heat treatment environment and temperature control between the two labs.

## V. Conclusion

The $I_c$ and RRR test results for the verification of the LARP wires are presented. A comparison between the supplier and the NHMFL is shown. Our results are in reasonable agreement with the supplier's quality control data. The reasons for the small but systematically higher $I_c$ and RRR values by NHMFL compared with those of the manufacturer are discussed.

## Acknowledgment

We would like to acknowledge Bruker OST for development and fabrication of the $Nb_3Sn$ wire for LARP program and for providing the quality control measurement data.